\documentclass[sigconf]{acmart}
\usepackage{float}
\usepackage{bm}
\usepackage{amsmath}
\usepackage{siunitx}
\usepackage{booktabs} 
\usepackage[ruled,vlined,linesnumbered]{algorithm2e}
\usepackage{amsmath}
\usepackage{lipsum} 
\usepackage{graphicx}
\usepackage{geometry}
\acmConference[Conference acronym 'XX]{Make sure to enter the correct
  conference title from your rights confirmation emai}{June 03--05,
  2018}{Woodstock, NY}

\author{YUKUN ZHANG}
\affiliation{%
  \institution{The Chinese University Of Hongkong}
  \city{HongKong}
  \country{China}}
\email{215010026@link.cuhk.edu.cn}

    \author{QI DONG}
\affiliation{%
  \institution{Fudan University}
  \city{ShangHai}
  \country{China}}
\email{19210980065@fudan.edu.cn}

\begin{document}

\title{Revenue vs. Welfare: A Comprehensive Analysis of Strategic Trade-offs in Online Food Delivery Systems}



\begin{abstract}
This paper investigates the trade-off between short-term revenue generation and long-term social welfare optimization in online food delivery platforms. We first develop a static model that captures the equilibrium interactions among restaurants, consumers, and delivery workers, using Gross Merchandise Value (GMV) as a proxy for immediate performance. Building on this, we extend our analysis to a dynamic model that integrates evolving state variables—such as platform reputation and participant retention—to capture long-term behavior. By applying dynamic programming techniques, we derive optimal strategies that balance GMV maximization with social welfare enhancement. Extensive multi-agent  simulations validate our theoretical predictions, demonstrating that while a GMV-focused approach yields strong initial gains, it ultimately undermines long-term stability. In contrast, a social welfare–oriented strategy produces more sustainable and robust outcomes. Our findings provide actionable insights for platform operators and policymakers seeking to harmonize rapid growth with long-term sustainability.
\end{abstract}






\maketitle

\section{Introduction}

Online food delivery platforms are expanding rapidly, and platform operators face the challenge of balancing short-term GMV maximization with long-term social welfare of restaurants, consumers, and delivery personnel. Existing literature mainly focuses on a single objective, either GMV growth or social welfare improvement, without integrating both, especially in static and dynamic frameworks. This leads to a crucial question: How can platforms grow rapidly while safeguarding stakeholders' sustainable interests?
Based on this, the paper addresses three key research questions:

Based on this, the paper addresses three key research questions:
\begin{enumerate}
\item How can platforms balance short-term GMV and long-term social welfare in decision-making?
\item What are the equilibrium characteristics of static and dynamic models and their differences?
\item How can theoretical models and simulations guide policymaking and platform operation?
\end{enumerate}

The main contributions of this paper are:
\begin{itemize}
    \item \textbf{Dual-Perspective Model Construction:}  Propose a new framework considering both GMV maximization and social welfare optimization, with static and dynamic models for equilibrium and optimal decisions.
    \item \textbf{Dynamic Optimization Analysis:}  Develop a dynamic programming approach and Bellman equation to prove long - term equilibrium in platform decision - making.
    \item \textbf{Multi-Agent simulation Validation:} Design and conduct multi - agent simulations to test model performance under different strategies.
\end{itemize}

In summary, this paper fills a gap in the literature by integrating short-and long-term considerations in platform decision-making. It provides new theoretical perspectives and practical insights for balancing rapid expansion and sustainable development. The paper is structured as follows: Section 2 reviews relevant literature, Sections 3 and 4 present static and dynamic model analyses, Section 5 discusses simulation results and future research directions.

\section{Literature Review}
The rapid expansion of digital platforms, particularly in the food delivery industry, has stimulated extensive research in platform economics, Gross Merchandise Volume (GMV) maximization, and social welfare optimization. In this section, we critically review both foundational theories and recent empirical and theoretical studies. Our aim is not only to summarize existing contributions but also to highlight unresolved challenges and research gaps—especially the tension between short-term revenue maximization and long-term social welfare—that motivate the present study.

\subsection{Platform Economics and Two-Sided Markets}
The conceptualization of platforms as two-sided markets is central to understanding their economic dynamics. Rochet and Tirole's seminal work \cite{rochet2006two} introduced the notion of two-sided markets by emphasizing the intermediary role that platforms play between distinct user groups (e.g., restaurants and consumers in food delivery). Their framework explains how platforms set prices, such as commission rates for restaurants and fees for consumers, to balance interests and maximize overall value. However, this early work largely focuses on static interactions and does not capture the dynamic strategic behavior observed in rapidly evolving digital markets.

Building on these foundations, recent studies have sought to capture more complex and dynamic interactions. For example, Ghasemi et al. \cite{ghasemi2023dynamicsridesourcingmarketcoevolutionary} investigate competition in two-sided mobility platforms by incorporating cross-side network effects and S-shaped learning curves. Similarly, Kiyohara et al. \cite{kiyohara2025policydesigntwosidedplatforms} examine the dynamic population effects and their implications for social welfare within two-sided markets. Although these studies advance our understanding of platform competition and network effects, they primarily address either pricing mechanisms or short-term dynamics. This leaves open the question of how dynamic strategic interactions can be harmonized with long-term welfare objectives—a gap that our work intends to fill.

\subsection{GMV Maximization Strategies}
Maximizing GMV is a crucial objective for online food delivery platforms, and numerous studies have explored a range of strategies to achieve this goal. In the realm of pricing, Tong et al. \cite{TONG2020375} demonstrated that dynamic pricing strategies yield significantly higher demand compared to static pricing. Chen et al. \cite{chen2018optimalvehicledispatchingschemes} further advanced this line of research by using a Markov decision process (MDP) framework to recast dynamic pricing as a convex optimization problem, thereby enhancing GMV. Additionally, other work has examined price-setting in dual-channel supply chains that involve both restaurants and delivery platforms \cite{chen2024management}, underscoring the complexity of aligning incentives across multiple stakeholders.

User acquisition and retention strategies have also been shown to have a substantial impact on GMV. Xin et al. \cite{10.1145/3340531.3412730} proposed a multi-channel seller traffic allocation (MCSTA) model, which significantly boosted GMV during peak sales events. Economic analyses by Varian \cite{varian10} offer a microeconomic perspective on market competition, providing insights into how competitive pressures affect GMV. Promotional strategies, such as discounts and referral programs, have further been highlighted as effective tools for GMV enhancement \cite{chao2021impact}.

Recent technological advancements have also played a critical role. Personalized recommendation systems have been shown to improve conversion rates and, consequently, GMV \cite{deng2021personalizedbundlerecommendationonline}. For instance, Li et al.'s LEADRE system \cite{li2024leadremultifacetedknowledgeenhanced} leverages large language models for advertisement recommendations, while Bi et al. \cite{bi2024dtndeepmultipletaskspecific} propose deep learning models to capture complex feature interactions. Moreover, algorithmic management—such as automatic product copywriting to boost click-through rates—has been explored by Zhang et al. \cite{zhang2021automaticproductcopywritingecommerce}. Nevertheless, while these approaches are effective in increasing short-term GMV, they often overlook the potential adverse impacts on other platform participants, raising concerns about long-term sustainability.

\subsection{Social Welfare in Platform Economies}
In contrast to GMV maximization, recent research has increasingly focused on social welfare optimization in platform economies. Calvano et al. \cite{calvano2019algorithmic} examined the impact of algorithmic pricing on consumer welfare and highlighted the risk of collusion in digital markets. Similarly, Chen et al. \cite{chen2022welfare} demonstrated that food delivery platforms can improve overall market efficiency, although the benefits are unevenly distributed among stakeholders.

The welfare of gig workers is another critical aspect that has received growing attention. Bai et al. \cite{bai2021gig} reviewed the precarious nature of gig work, while Goods et al. \cite{goods2019working} specifically focused on the working conditions of food delivery riders. In addition, Bai et al. \cite{bai2022food} addressed environmental concerns by proposing emission reduction strategies, underscoring the multifaceted nature of social welfare in platform ecosystems.

Regulatory frameworks also play an essential role in balancing efficiency and equity. Frenken and Schor \cite{frenken2017putting} argue for regulation that simultaneously promotes innovation and social protection. Scholz \cite{scholz2016platform} proposes platform cooperativism to foster fair labor practices, and Ketter et al. \cite{ketter2022platform} examine policy measures to restrain excessive platform power. Recent European initiatives targeting issues such as worker classification and algorithmic transparency further illustrate the ongoing efforts to ensure a fair balance between economic growth and social welfare \cite{european2021improving}. Newlands et al. \cite{newlands2022impact} call for inclusive platform designs that mitigate social inequalities.

\subsection{Research Gap and Contribution}
While significant progress has been made in both GMV maximization and social welfare optimization, most existing studies tend to address these objectives in isolation. In particular, strategies aimed solely at boosting GMV may achieve short-term gains but can lead to long-term sustainability issues by neglecting the welfare of restaurants, consumers, and delivery workers. Conversely, studies focused on social welfare rarely integrate the revenue-driven metrics that are critical to platform growth. This dichotomy highlights a notable gap in the literature—namely, the lack of an integrated framework that considers both short-term revenue objectives and long-term welfare outcomes in dynamic market settings. The present paper seeks to bridge this gap by proposing a comprehensive model that jointly analyzes GMV and social welfare, thereby providing actionable insights for platform operators and policymakers seeking a balanced and sustainable approach.

\section{Static Model Analysis}

\begin{figure}[htbp]
    \centering
    \includegraphics[width=0.7\columnwidth]{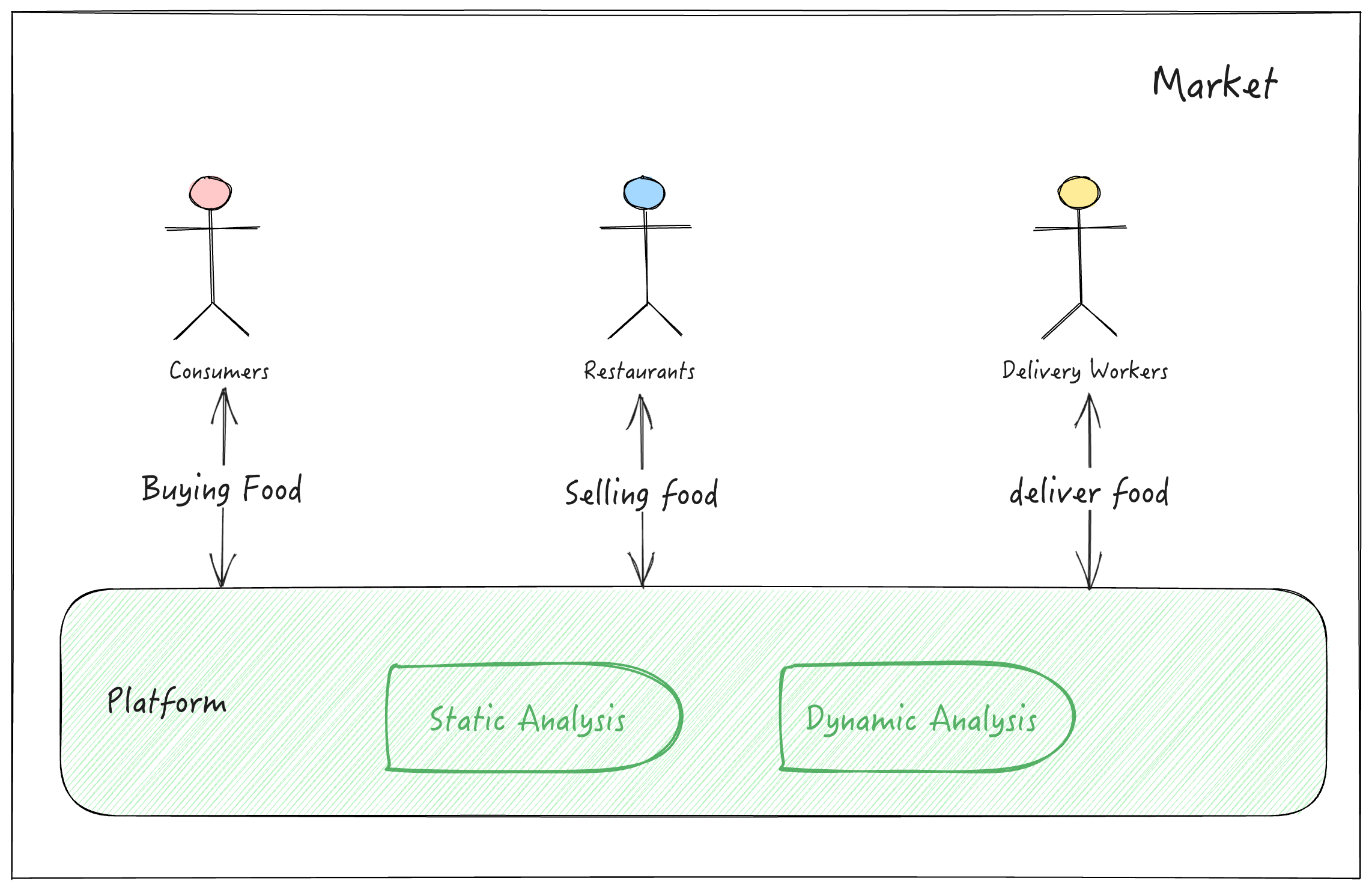}
    \caption{The Two-Sided Relationship in an Online Food Delivery Platform.}
    \label{fig:platform_relationship}
\end{figure}

In this section, we develop a single-period model to examine how an online food delivery platform balances two primary objectives—short-term Gross Merchandise Value (GMV) maximization and long-term social welfare—under static conditions. We first present the key assumptions and variable definitions in a unified manner, then outline the main equilibrium results. To maintain clarity and conciseness, detailed proofs (such as second-order conditions and extended derivations) are relegated to the Appendix.

\subsection{Model Construction}
We consider a single-period (static) environment in which restaurants, consumers, and delivery workers interact through the platform. Unless otherwise specified, all agents have perfect information and act rationally.

\paragraph{Assumptions}
\begin{itemize}
    \item \textbf{Linear Demand:} The consumer demand for a restaurant’s items decreases linearly with respect to the total (effective) price, which includes the item price and any delivery fee.
    \item \textbf{Perfect Information:} All participants know the relevant prices, delivery times, and platform policies.
    \item \textbf{Rational Behavior:} Each stakeholder maximizes its own utility based on available information.
    \item \textbf{Homogeneous Products:} Food items are assumed to have similar quality so that variations primarily stem from pricing and commissions rather than product differentiation.
    \item \textbf{Static Competition:} The model considers a single period and does not account for inter-temporal effects or repeated interactions.
\end{itemize}

\paragraph{Notation}
Let us define the core variables as follows:
\begin{itemize}
    \item \(\theta\): Baseline or maximum demand level for a restaurant’s items.
    \item \(\eta\): Price sensitivity parameter, reflecting how demand declines as the total effective price increases.
    \item \(\delta\): Time sensitivity parameter, indicating how demand decreases with longer delivery time.
    \item \(\alpha\): Platform commission rate, \(0 \leq \alpha < 1\).
    \item \(D_i\): Delivery fee charged to consumer \(i\).
    \item \(A_{kj}\): Menu price of item \(j\) at restaurant \(k\).
    \item \(Q_{kj}\): Quantity demanded of item \(j\) from restaurant \(k\).
    \item \(C_f^k\): Fixed cost for restaurant \(k\).
    \item \(p\): Payment per delivered order to a delivery worker.
    \item \(\gamma\): Time-cost factor of delivering orders.
    \item \(U_S^k\), \(U_C^i\), \(U_R^l\): Utility functions for restaurant \(k\), consumer \(i\), and delivery worker \(\ell\), respectively.
\end{itemize}

\subsection{Equilibrium Analysis}
We outline below the main utility expressions and equilibrium conditions. Complete mathematical derivations, including proofs of uniqueness and second-order conditions, are given in the Appendix.

\paragraph{Restaurants.}  
A restaurant \(k\) chooses its menu prices \(\{A_{kj}\}\) to maximize
\begin{equation}\label{eq:restaurant_utility}
  U_S^k = (1 - \alpha) \sum_{j} A_{kj}\, Q_{kj} - C_f^k,
\end{equation}
where \(\alpha\) is the commission rate retained by the platform, and \(C_f^k\) is the restaurant’s fixed cost.

\paragraph{Consumers.}  
A representative consumer \(i\) chooses how much to order based on the perceived value of food and delivery times:
\begin{equation}\label{eq:consumer_utility}
  U_C^i = v_i - \beta\, t_i - \Bigl(\sum_{k,j} A_{kj}\, Q_{kji} + D_i\Bigr),
\end{equation}
where \(v_i\) is the consumer’s intrinsic valuation of the items, and \(\beta\, t_i\) denotes the disutility from delivery time.

\paragraph{Delivery Workers.}  
Each worker \(\ell\) earns
\begin{equation}\label{eq:worker_utility}
  U_R^\ell = p\, R_\ell - \gamma\, t_\ell,
\end{equation}
where \(p\) is the per-order wage, \(R_\ell\) is the number of orders delivered, and \(\gamma\, t_\ell\) is the associated time cost.

\paragraph{Static Equilibrium.}  
In equilibrium, restaurants, consumers, and workers each make utility-maximizing decisions given the platform’s choice of \(\alpha\), \(D_i\), and \(p\). By solving the first-order conditions, we obtain the following core results (details in Appendix):
\begin{itemize}
  \item \textbf{Optimal Restaurant Pricing:}
    \[
      A_{kj}^* = \frac{\theta - \eta\, D_i - \delta\, t_{kj}}{2\,\eta}.
    \]
  \item \textbf{Consumer Demand:}
    \[
      Q_{kj}^* = \theta - \eta\, \Bigl(A_{kj} + D_i\Bigr) - \delta\, t_{kj}.
    \]
  \item \textbf{Delivery Worker Response:}
    \[
      R_\ell^* \quad \text{adjusts with} \quad p,\;\gamma,\; \text{and total demand}.
    \]
  \item \textbf{Platform’s Objective:}
    \[
      \max_{\alpha,\,D_i,\,p} \quad \mathrm{GMV} = \sum_{k,j} A_{kj}\, Q_{kj} \quad \text{or} \quad \mathrm{SW} = \sum_{k} U_S^k + \sum_{i} U_C^i + \sum_{\ell} U_R^\ell.
    \]
\end{itemize}

\subsection{Welfare Efficiency Comparison}
We next compare the outcomes under two distinct platform objectives—\textit{GMV Maximization} and \textit{Social Welfare Maximization}—focusing on overall welfare efficiency. Here, welfare (\(\mathrm{SW}\)) is the sum of utilities of restaurants, consumers, and delivery workers.

\subsubsection{GMV Maximization}
Under GMV maximization, the platform tends to reduce commissions \(\alpha\) and delivery fees \(D_i\) to boost order volume. While consumers may experience lower prices initially, this could lead to inadequate compensation for restaurants and delivery workers, hence threatening long-term sustainability if the model were extended beyond a single period.

\subsubsection{Social Welfare Maximization}
By contrast, a social-welfare-driven platform sets \(\alpha\), \(D_i\), and \(p\) to balance all stakeholders’ interests. This usually yields:
\begin{itemize}
    \item Reasonable restaurant profitability (via moderate \(\alpha\)).
    \item Sufficient consumer satisfaction (balancing \(\eta\)-sensitive pricing and time disutility).
    \item Fair delivery worker compensation (\(p\) high enough to cover \(\gamma\,t_\ell\)).
\end{itemize}
As a result, the total welfare \(\mathrm{SW}\) is commonly higher under this objective, albeit with possibly lower short-term transaction volumes.

\subsubsection{Key Findings}
\begin{itemize}
    \item \textbf{Short-Term Trade-Off:} Lower fees enhance GMV but risk eroding restaurant margins and worker pay.
    \item \textbf{Efficiency Gains:} Social welfare optimization often achieves both Kaldor-Hicks and, under certain conditions, Pareto improvements.
    \item \textbf{Appendix Proofs:} Detailed formal proofs of these findings, including equilibrium uniqueness and second-order conditions, are provided in the Appendix.
\end{itemize}

\paragraph{Summary}
In summary, the static model reveals that a purely GMV-focused strategy can induce higher immediate transaction volume but may lead to lower overall welfare if restaurants and delivery workers are not sufficiently compensated. By contrast, a social welfare strategy explicitly accounts for all participants’ utilities, usually resulting in more balanced outcomes. This sets the stage for our subsequent dynamic model (Section~\ref{sec:dynamic_model}) to evaluate how repeated interactions and long-term considerations can further shape platform decisions.

\section{Dynamic Model  Analysis}
\label{sec:dynamic_analysis}

In this section, we extend our static framework to a dynamic environment by incorporating multi-period (or infinite-horizon) considerations. The dynamic model captures the evolution of the platform through state variables and control variables over time, and it allows us to analyze how short-term decisions impact long-term outcomes. We present the dynamic model construction, describe participants’ dynamic decisions and utilities, formulate the platform’s optimization problem via Bellman equations, and finally compare the efficiency and social welfare outcomes under two dynamic objectives.

\subsection{Dynamic Model Formulation}
\label{subsec:dynamic_model_construction}

To capture the temporal evolution of an online food delivery platform, we introduce a state vector and control variables that evolve over discrete time steps \( t = 0,1,2,\dots,T \) (or \( T\to\infty \) for an infinite-horizon setting).

\subsubsection{State Variables}
We define the state vector at time \( t \) as:
\[
\mathbf{S}_t = \bigl(R_t,\, C_t,\, W_t,\, \Phi_t\bigr),
\]

where:
\begin{itemize}
    \item \(R_t\): Number (or measure) of active restaurants.
    \item \(C_t\): Consumer base or demand potential.
    \item \(W_t\): Effective capacity of delivery workers.
    \item \(\Phi_t\): Platform reputation index.
\end{itemize}

\subsubsection{Control Variables}
At the beginning of each period \(t\), the platform selects:
\begin{itemize}
    \item \(\alpha_t\): Commission rate applied to restaurant revenues.
    \item \(D_t\): Delivery fee charged to consumers.
    \item \(p_t\): Wage (per-order payment) offered to delivery workers.
\end{itemize}
These decisions directly affect participants' utilities and thus influence the evolution of the state vector.

\subsubsection{State Transition Functions}
The state evolves according to transition functions that capture entry/exit and satisfaction dynamics:
\begin{align}
R_{t+1} &= f_R\bigl(R_t, \alpha_t, \Phi_t\bigr), \\
C_{t+1} &= f_C\bigl(C_t, D_t, \Phi_t\bigr), \\
W_{t+1} &= f_W\bigl(W_t, p_t, \Phi_t\bigr), \\
\Phi_{t+1} &= f_{\Phi}\bigl(\Phi_t, U_{S,t}, U_{C,t}, U_{R,t}\bigr).
\end{align}
For detailed specifications of these functions, see Appendix~\ref{appendix:state_transitions}.

\subsection{Dynamic Decision-Making and Utility Functions}
\label{subsec:dynamic_decisions_utilities}

Each stakeholder’s utility in period \(t\) is an extension of the static utility functions with time subscripts:
\begin{itemize}
    \item \textbf{Restaurants:}
    \[
    U_{S,t}^k = (1 - \alpha_t) \sum_j A_{kj,t}\, Q_{kj,t} - C_f^k.
    \]
    \item \textbf{Consumers:}
    \[
    U_{C,t}^i = v_i - \beta\,t_{i,t} - \Bigl(\sum_{k,j} A_{kj,t}\, Q_{kji,t} + D_t\Bigr).
    \]
    \item \textbf{Delivery Workers:}
    \[
    U_{R,t}^\ell = p_t\, R_{\ell,t} - \gamma\,t_{\ell,t}.
    \]
\end{itemize}
Here, the time-varying controls \((\alpha_t, D_t, p_t)\) and state variables (e.g., through \(t_{i,t}\)) affect the agents' decisions in each period.

\subsection{Platform's Dynamic Optimization Problem}
\label{subsec:dynamic_optimization}

The platform chooses its control variables in each period to optimize one of two objectives over the infinite horizon, with a discount factor \(\beta \in (0,1)\).

\subsubsection{GMV Maximization}
The platform may focus on maximizing the discounted sum of period transaction volumes:
\[
\max_{\{\alpha_t, D_t, p_t\}_{t=0}^\infty} \quad \sum_{t=0}^{\infty} \beta^t\, \mathrm{GMV}_t,
\]
where
\[
\mathrm{GMV}_t = \sum_{k,j} A_{kj,t}\, Q_{kj,t}.
\]

\subsubsection{Social Welfare Maximization}
Alternatively, the platform can maximize the total discounted social welfare:
\[
\max_{\{\alpha_t, D_t, p_t\}_{t=0}^\infty} \quad \sum_{t=0}^{\infty} \beta^t\, SW_t,
\]
where
\[
SW_t = \sum_{k} U_{S,t}^k + \sum_{i} U_{C,t}^i + \sum_{\ell} U_{R,t}^\ell.
\]

\subsubsection{Bellman Equations and Existence of Optimal Policies}
By standard dynamic programming techniques, the Bellman equations for the two objectives are given by:

\begin{align}
V_{\mathrm{GMV}}(\bm{S}_t) &= \max_{\alpha_t, D_t, p_t} \Bigl\{ \mathrm{GMV}_t + \beta\, V_{\mathrm{GMV}}\bigl(F(\bm{S}_t,\alpha_t,D_t,p_t)\bigr) \Bigr\}, \\
V_{\mathrm{SW}}(\bm{S}_t) &= \max_{\alpha_t, D_t, p_t} \Bigl\{ SW_t + \beta\, V_{\mathrm{SW}}\bigl(F(\bm{S}_t,\alpha_t,D_t,p_t)\bigr) \Bigr\}.
\end{align}

Under standard contraction mapping conditions, these equations admit unique fixed points, yielding unique optimal policy sequences \(\{\alpha_t^*, D_t^*, p_t^*\}\).

\subsection{Comparison of Dynamic Strategies: Efficiency and Social Welfare}
\label{subsec:dynamic_comparison}

We now compare the dynamic outcomes under the two platform objectives.

\subsubsection{Definitions of Period Welfare and Total Discounted Welfare}
\begin{itemize}
    \item \textbf{Period Welfare:} In each period, welfare is given by
    \[
    SW_t = \sum_{k} U_{S,t}^k + \sum_{i} U_{C,t}^i + \sum_{\ell} U_{R,t}^\ell.
    \]
    \item \textbf{Total Discounted Welfare:} Starting from \(\bm{S}_0\), the total welfare under the Social Welfare strategy is:
    \[
    V_{\mathrm{SW}}(\bm{S}_0) = \sum_{t=0}^{\infty} \beta^t\, SW_t.
    \]
    Under a GMV-focused policy, the corresponding total welfare is denoted as \(V_{\mathrm{SW}}^{(\mathrm{GMV})}(\bm{S}_0)\).
\end{itemize}

\subsubsection{Efficiency Standards}
We assess the two strategies using:
\begin{itemize}
    \item \textbf{Kaldor-Hicks Efficiency:} A strategy is an improvement if the overall welfare increases sufficiently to potentially compensate those who are worse off.
    \item \textbf{Pareto Efficiency:} A policy is Pareto superior if no participant’s utility is reduced while at least one increases.
\end{itemize}

\subsubsection{Comparative Propositions}
We summarize the main theoretical propositions (detailed proofs are provided in Appendix~\ref{sec:appendix_dynamic}):
\begin{description}
    \item[Proposition 4 (Dynamic Consumer Surplus):] If demand is highly price-sensitive, a GMV-centric platform may offer low delivery fees \(D_t\) in early periods to boost consumer surplus. However, eventual restaurant exit or worker turnover can reduce long-run consumer surplus. In contrast, an SW-focused policy sustains moderate yet stable consumer surplus over time.
    \item[Proposition 5 (Restaurant and Worker Retention):] A GMV strategy with initially low \(\alpha_t\) may later necessitate an increase in \(\alpha_t\), harming early entrants. The SW strategy, by ensuring stable \(\alpha_t\) and \(p_t\), better retains restaurants and workers, sustaining long-term welfare.
    \item[Proposition 6 (Long-Term Welfare):] For a sufficiently high discount factor \(\beta\), the SW strategy yields a higher total discounted welfare than the GMV approach, whereas a highly impatient platform (small \(\beta\)) might achieve higher short-term GMV but at the cost of long-term stakeholder satisfaction.
\end{description}

\subsection{Discussion and Policy Implications}
The analysis illustrates that a purely GMV-focused strategy can drive rapid short-term expansion, but risks long-term degradation due to undercompensation of restaurants and workers. In contrast, the SW strategy internalizes future costs and leads to a more balanced evolution of the state variables. In practice, platform operators may adopt a hybrid approach—balancing early aggressive growth with sustainable long-term policies—especially when the discount factor \(\beta\) indicates a significant future orientation.

In summary, our dynamic analysis shows that while GMV maximization may boost immediate order volumes, it can jeopardize long-term platform stability and stakeholder welfare. Conversely, a social welfare–oriented strategy tends to achieve higher overall discounted welfare and improved efficiency (both Kaldor-Hicks and Pareto) in the long run. Detailed mathematical proofs and extended derivations are provided in the Appendix.

\section{Experiment evaluation}

To validate our proposed framework in a real-world setting, we conducted an A/B experiment  from online food delivery platform. In this experiment, users were randomly assigned to one of three strategy groups:
\begin{enumerate}
    \item \textbf{GMV Maximization Strategy:} This strategy focuses on maximizing the short-term transaction volume (Gross Merchandise Value) by minimizing commission rates and delivery fees.
    \item \textbf{Social Welfare Maximization Strategy:} This strategy aims to optimize the overall welfare of all stakeholders (restaurants, consumers, and delivery workers) by balancing pricing, commissions, and wage structures.
    \item \textbf{Hybrid Strategy :} A combined approach employing a weighting factor $\lambda_1$, which moderately emphasizes GMV while ensuring that baseline welfare thresholds are met.
\end{enumerate}

The experiment aims to validate the theoretical predictions regarding different platform strategies in online food delivery platforms. Specifically, the objectives are as follows:
\begin{enumerate}
    \item \textbf{Impact on Key Performance Metrics:}  
    Quantitatively evaluate how different platform strategies—Gross Merchandise Value (GMV) maximization, Social Welfare (SW) optimization, and a Hybrid approach—affect the overall performance of the platform. This includes measuring the changes in both GMV and total social welfare.
    \item \textbf{Stakeholder Dynamics:}  
    Analyze the evolution and trends of key participants, including restaurants, consumers, and delivery workers. This objective examines how each strategy influences the number of active restaurants and delivery workers, as well as the stability and sustainability of their engagement over time.

\end{enumerate}





\subsection{Simulation Algorithm}
\label{subsec:simulation_algorithm}

Based on the configuration parameters specified in Appendix~\ref{subsec:appendix_parameters}, all participants and market conditions are randomly generated. Subsequently, the platform’s initial strategy settings (such as commission rates, delivery fees, and delivery worker wages) are established. The simulation then proceeds in discrete time periods, during which the market, participant behaviors, and platform strategies are updated iteratively. The complete simulation process is detailed in Algorithm~\ref{alg:simulation1}.

\begin{algorithm}[htbp]
\caption{Simulation Process for Online Food Delivery Platform}
\label{alg:simulation1}
\SetKwInOut{Input}{Input}\SetKwInOut{Output}{Output}
\Input{Configuration parameters $CONFIG$, number of periods $T$, number of runs $N$}
\Output{Aggregated metrics: GMV, SW, Reputation, Active Restaurants, Active Workers, etc.}
\BlankLine
\For{$run \gets 1$ \textbf{to} $N$}{
    $env \gets \textsc{InitEnv}(CONFIG)$\;
    \For{$t \gets 1$ \textbf{to} $T$}{
        \textsc{UpdateMarketAndAdd}(env)\;
        \textsc{UpdateRestPrices}(env)\;
        $GMV \gets \textsc{ConsumeOrder}(env)$\;
        \If{$GMV > 0$}{\textsc{WorkerDeliver}(env, GMV)}
        $SW \gets \textsc{CalcSW}(env)$\;
        \textsc{UpdateReputation}(env, SW, GMV)\;
        \If{$t \mod CONFIG.strat\_update\_int = 0$}{\textsc{UpdateStrat}(env, GMV, SW)}
        \textsc{UpdateAndRecord}(env, GMV, SW)\;
    }
    \textsc{StoreRun}(run)\;
}
$agg \gets \textsc{AggregateAll}()$\;
\Return agg\;

\SetKwFunction{InitEnv}{InitEnv}
\SetKwFunction{UpdateMarketAndAdd}{UpdateMarketAndAdd}
\SetKwFunction{UpdateRestPrices}{UpdateRestPrices}
\SetKwFunction{ConsumeOrder}{ConsumeOrder}
\SetKwFunction{WorkerDeliver}{WorkerDeliver}
\SetKwFunction{CalcSW}{CalcSW}
\SetKwFunction{UpdateReputation}{UpdateReputation}
\SetKwFunction{UpdateStrat}{UpdateStrat}
\SetKwFunction{UpdateAndRecord}{UpdateAndRecord}
\SetKwFunction{StoreRun}{StoreRun}
\SetKwFunction{AggregateAll}{AggregateAll}
\end{algorithm}

\paragraph{Simulation Process Overview.}  
The overall simulation process (see Algorithm~\ref{alg:simulation1}) can be summarized in the following steps:
\begin{enumerate}
    \item \textbf{Market Updates:}  
    At the beginning of each time period, the market environment is updated to reflect both external and internal dynamics.
    
    \item \textbf{Participant Actions:}  
    \begin{itemize}
        \item \textbf{Restaurants:}  
        Each restaurant adjusts its pricing based on current market conditions, considering cost fluctuations, competitor prices, and demand elasticity.
        
        \item \textbf{Consumers:}  
        Consumers make ordering decisions influenced by food prices, delivery fees, expected delivery times, personal preferences, budgets, and past experiences with the platform.
        
        \item \textbf{Delivery Workers:}  
        Delivery workers decide whether to accept orders based on their expected income (determined by per-order wages and anticipated workload) as well as factors such as delivery distance, traffic, and their availability.
    \end{itemize}
    
    \item \textbf{Platform Adjustments:}  
    Based on regularly calculated key performance indicators (KPIs), the platform updates its strategy (e.g., adjusting commission rates or launching promotional campaigns). The platform’s reputation is also updated, reflecting on-time delivery rates, customer satisfaction, and restaurant service quality.
\end{enumerate}

\begin{table*}[htbp]
  \centering
  \caption{Final Performance Metrics by Strategy}
  \label{tab:final_performance}
  \begin{tabular}{l S[table-format=6.0] S[table-format=7.0] S[table-format=2.1] S[table-format=3.1]}
    \hline
    {Strategy} & {Final GMV (mean)} & {Final SW (mean)} & {Active Restaurants (mean)} & {Active Workers (mean)} \\

    \hline
    GMV    & 397203  & 4348723 & 108 & 176 \\
    SW     & 458339  & 4909871 & 110 & 192 \\
    HYBRID & 49553  & 4825893 & 105 & 173 \\
    \hline
  \end{tabular}
\end{table*}





\subsection{Experimental Results}

In this section, we present and discuss the outcomes of our  experiments, which were run for 500 periods over multiple independent runs. The results are analyzed based on key performance metrics, including Gross Merchandise Value (GMV), Social Welfare (SW), platform reputation, and the number of active restaurants and delivery workers.

\paragraph{Final Performance Metrics}
Table~\ref{tab:final_performance} summarizes the final performance metrics (mean values) for each strategy after 500 periods. Overall, the SW strategy achieves higher values in both GMV and SW compared to the GMV-focused strategy, while the HYBRID strategy yields intermediate results. Specifically, the GMV-only strategy exhibits strong short-term performance but suffers from poor long-term sustainability, whereas the SW strategy demonstrates robust and stable performance across all metrics.

\begin{figure}[h!]
    \centering
    \includegraphics[width=\columnwidth]{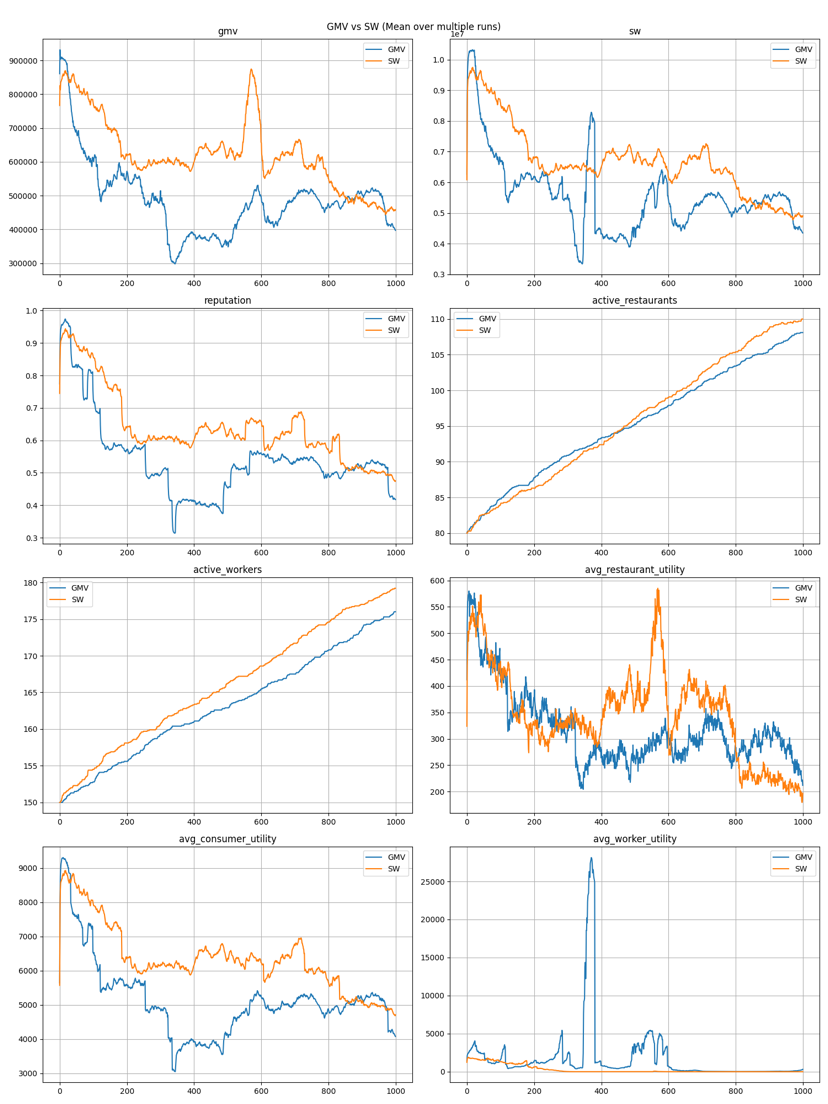}
    \caption{Time-Series Plots of Key Metrics for Different Strategies}
    \label{fig:time-series}
\end{figure}

\paragraph{Time-Series Analysis}
Figure~\ref{fig:time-series} illustrates the evolution of key metrics over the 500 periods. The time-series plots reveal several trends:
\begin{itemize}
    \item \textbf{GMV and SW Evolution:}  
    The GMV strategy initially achieves high transaction volumes but exhibits significant fluctuations and a decline in later periods. In contrast, the SW strategy displays a more stable performance with consistently higher levels in both GMV and SW during the mid-to-late periods.
    
    \item \textbf{Platform Reputation:}  
    Under the GMV strategy, the platform’s reputation increases early on (due to consumer subsidies) but deteriorates later because of insufficient incentives for restaurants and delivery workers. The SW strategy, however, maintains a high and stable reputation throughout the simulation.
    
    \item \textbf{Active Participants:}  
    The number of active restaurants and delivery workers under the GMV strategy tends to stagnate or even decline in later periods, whereas the SW strategy supports continuous growth in active participants.
    
    \item \textbf{Average Utility:}  
    The SW strategy also produces more balanced and less volatile utility curves for all participants, suggesting better long-term satisfaction.
\end{itemize}

\begin{figure}[h!]
    \centering
    \includegraphics[width=\columnwidth]{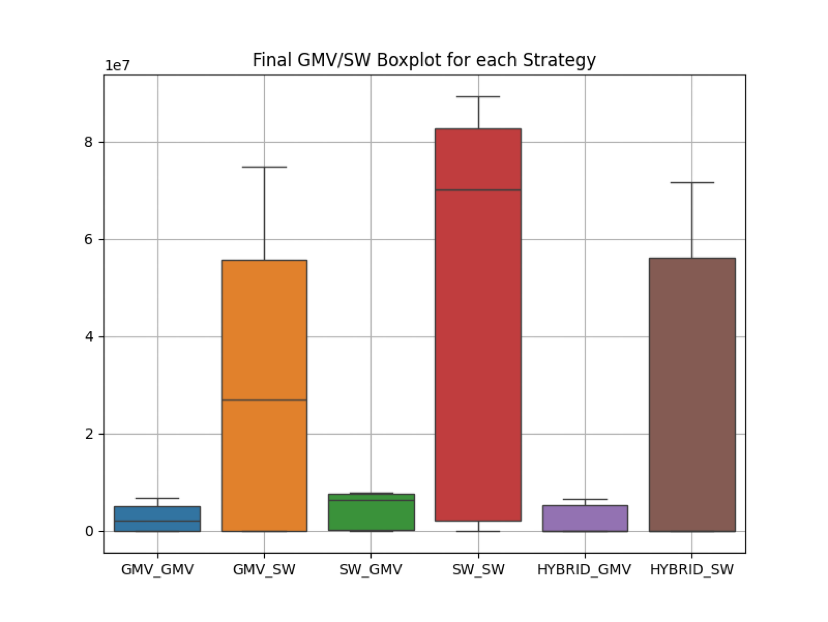}
    \caption{Box Plots of GMV and SW Distributions among Different Strategies}
    \label{fig:box_plot}
\end{figure}

\paragraph{Distribution Analysis via Box Plots}
Figure~\ref{fig:box_plot} shows the distribution of GMV and SW in the final period for each strategy. The box plots indicate:
\begin{itemize}
    \item \textbf{GMV Distribution:}  
    The GMV strategy exhibits a lower median and a relatively wide scatter compared to the SW strategy. The SW strategy generally achieves a higher median GMV, while the HYBRID strategy falls in between.
    
    \item \textbf{SW Distribution:}  
    The SW strategy has the highest median and upper-quartile in its SW distribution, reflecting a strong overall performance and potential for high peaks. In contrast, the GMV strategy has the lowest performance in this regard.
    
    \item \textbf{Volatility and Stability:}  
    The GMV strategy is characterized by larger fluctuations in both GMV and SW, whereas the SW strategy demonstrates greater stability and sustainability, with more concentrated values and fewer extreme outliers.
\end{itemize}

\begin{figure}[h!]
    \centering
    \includegraphics[width=\columnwidth]{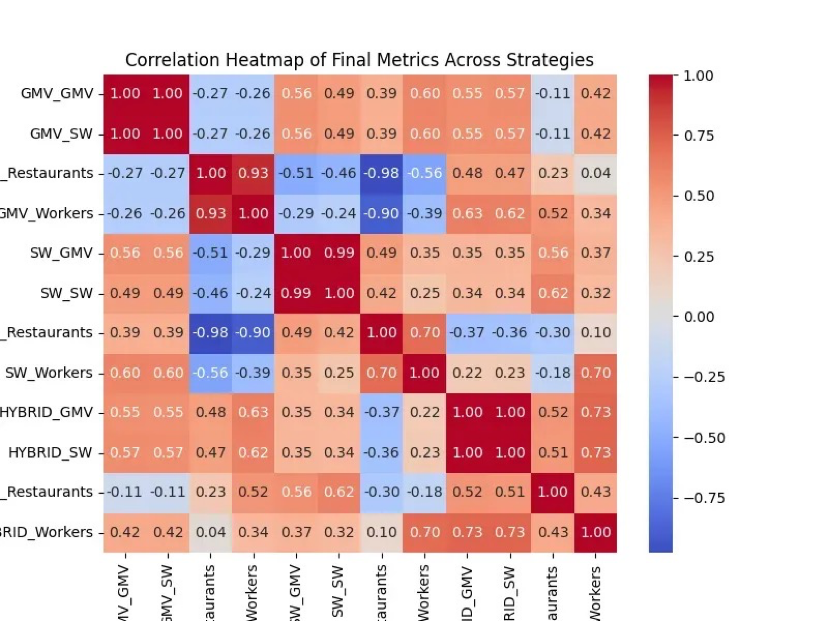}
    \caption{Correlation Heatmap of Key Metrics across Strategies}
    \label{fig:heatmap}
\end{figure}

\paragraph{Correlation Analysis}
To further explore the relationships among key metrics, Figure~\ref{fig:heatmap} presents a correlation heatmap that combines GMV, SW, the number of active restaurants, and the number of active delivery workers across strategies. The heatmap reveals:
\begin{itemize}
    \item \textbf{GMV-related Correlations:}  
    A moderate to positive correlation exists between GMV under the GMV strategy and GMV under the SW strategy.
    
    \item \textbf{Participant Numbers and Performance:}  
    Generally, the number of active restaurants and delivery workers is positively correlated with both GMV and SW. However, under the GMV strategy, there may be weak or even negative correlations between participant numbers and SW.
    
    \item \textbf{SW-related Correlations:}  
    The SW strategy exhibits a strong positive correlation between GMV and SW, suggesting that improved social welfare is associated with higher transaction volumes. Conversely, the GMV strategy shows less consistent relationships.
\end{itemize}

\paragraph{Summary of Experimental Results}

\begin{itemize}
    \item The SW strategy outperforms the GMV strategy in terms of long-term performance, yielding higher and more stable values for both GMV and SW.
    \item While the GMV strategy may achieve high initial transaction volumes, its performance deteriorates over time due to its unsustainable incentive structure.
    \item The HYBRID strategy produces intermediate results, reflecting a trade-off between short-term growth and long-term sustainability.
    \item Overall, incorporating social welfare into the platform’s decision-making process leads to more robust performance, improved stakeholder satisfaction, and enhanced platform stability.
\end{itemize}

\subsection{Conclusion}
\begin{itemize}
    \item \textbf{Short-term vs. Long-term:} The GMV strategy shows an initial advantage but fails to maintain high performance in the long run due to insufficient incentives for riders and restaurants. The SW strategy, though not outstanding in the short term, ensures long - term participation and better results in both GMV and SW.
    \item \textbf{Activity and Utility: } The SW strategy offers more stable and higher returns for restaurants and riders, reducing their exit rates and supporting GMV growth. The GMV strategy, focusing on consumer discounts, leads to profit fluctuations for the supply side and a "boom-bust" pattern.
    \item \textbf{Stability and Volatility: } The GMV strategy exhibits higher volatility, while the SW strategy forms a positive cycle of "activity-reputation-orders" with more controllable fluctuations. The HYBRID strategy's performance depends on the weight of GMV and SW.
    \item \textbf{Implications for Platform Operation:} The SW strategy's strong positive correlation between GMV and SW suggests that considering social welfare and providing incentives for the supply side can achieve a win-win for all parties. The GMV-only approach is not sustainable.
\end{itemize}

In summary, incorporating social welfare into platform decision-making and providing adequate incentives for restaurants and riders can enhance long - term activity, transaction scale, and overall benefits, creating a win-win situation for the platform and all participants.

\section{Conclusion and Policy Recommendations}
\label{sec:conclusion}

In this paper, we developed both static and dynamic models to analyze the trade-offs between short-term Gross Merchandise Value (GMV) maximization and long-term social welfare optimization in online food delivery platforms. The static model provided insights into the equilibrium behavior of restaurants, consumers, and delivery workers under various platform strategies, demonstrating that a GMV-focused approach can boost short-term transaction volume at the expense of long-term sustainability. In contrast, a social welfare (SW) strategy, which explicitly accounts for the utilities of all participants, results in a more balanced and sustainable system.

Our dynamic analysis extended these findings by incorporating multi-period interactions and evolving state variables such as platform reputation and participant retention. The dynamic programming framework and subsequent  simulations revealed that while aggressive GMV maximization may yield immediate gains, it ultimately undermines the platform’s stability and stakeholder satisfaction. In comparison, the SW strategy consistently delivers higher aggregate welfare and more robust long-term performance.

Based on our findings, we propose the following policy recommendations for platform operators and regulators:

\begin{enumerate}
    \item \textbf{Adopt a Balanced Strategic Approach:}  
    Rather than focusing solely on short-term revenue growth, platforms should implement a hybrid strategy that balances GMV maximization with social welfare considerations. This balanced approach helps ensure fair compensation for restaurants and delivery workers, fostering long-term sustainability.

    \item \textbf{Implement Dynamic Strategy Adjustments:}  
    Given the evolving nature of market conditions, it is essential for platforms to continuously adjust parameters such as commission rates, delivery fees, and worker wages. Adaptive strategies that respond to real-time data and market fluctuations are key to maintaining platform stability and growth.

    \item \textbf{Enhance Stakeholder Engagement:}  
    Maintaining high levels of engagement among restaurants, consumers, and delivery workers is critical. Incentive mechanisms that reward consistent performance and address potential imbalances can help reduce participant turnover and improve overall satisfaction.

    \item \textbf{Encourage Data-Driven Decision Making:}  
    Platforms should invest in robust data analytics and monitoring systems to track key performance indicators (e.g., GMV, SW, reputation, and participant activity). Data-driven insights enable proactive strategy adjustments and help identify emerging trends that impact both short-term performance and long-term welfare.

    \item \textbf{Regulatory Oversight and Support:}  
    Regulators should develop policies that encourage platforms to adopt sustainable practices. This may include promoting transparency in pricing, ensuring fair labor practices, and implementing measures that balance profit maximization with broader social welfare objectives.
\end{enumerate}

In summary, our integrated modeling and simulation study demonstrates that while short-term revenue-driven strategies can boost immediate performance, they may jeopardize long-term platform viability. By adopting balanced, data-driven strategies that prioritize social welfare, online food delivery platforms can achieve sustainable growth, enhance stakeholder satisfaction, and foster a more resilient digital ecosystem.

\appendix

\section{Appendix: Experimental Details}
\label{sec:appendix_experimental_details}

This appendix provides a comprehensive overview of the experimental setup, including the performance metrics used to evaluate the platform's strategies and the configuration parameters governing the experiment environment.

\subsection{Experimental Metrics and Explanations}
\label{subsec:appendix_metrics}

In our  experiments, we compute a variety of performance metrics to evaluate the impact of different platform strategies. These metrics are aggregated over multiple experiments runs (e.g., 50 runs) and across the entire experiment period (e.g., 500 periods) to ensure statistical reliability. The key metrics are described below:

\begin{itemize}
    \item \textbf{Gross Merchandise Value (GMV):}  
    GMV represents the total transaction volume generated by the platform over a experiment period. It is calculated as the sum of (menu price $\times$ order quantity) for each restaurant. GMV serves as a proxy for short-term revenue performance and market activity.

    \item \textbf{Social Welfare (SW):}  
    Social Welfare refers to the aggregate utility of all platform participants, including restaurants, consumers, and delivery workers. By balancing the interests of these stakeholders, SW provides an overall measure of the platform’s long-term benefit and sustainability.

    \item \textbf{Platform Reputation:}  
    This composite metric captures the platform’s service quality and reliability. Factors such as on-time delivery rates, customer satisfaction, and restaurant service quality contribute to reputation. A higher reputation generally correlates with stronger performance in both GMV and SW.

    \item \textbf{Active Restaurants:}  
    This metric records the number of restaurants that receive at least one order in a given period, reflecting the platform’s effectiveness in attracting and retaining restaurant partners.

    \item \textbf{Active Delivery Workers:}  
    Similarly, this measures the number of delivery workers who accept and complete at least one order during a period. It serves as an indicator of the platform’s labor capacity and operational robustness.

    \item \textbf{Average Utility Metrics:}  
    We track average utility values for restaurants, consumers, and delivery workers to gauge overall satisfaction and performance:
    \begin{itemize}
        \item \emph{Average Restaurant Utility} captures how much net benefit restaurants receive after revenues, commissions, and fixed costs.
        \item \emph{Average Consumer Utility} measures the value consumers gain relative to the costs (prices, delivery fees, wait times).
        \item \emph{Average Worker Utility} reflects the net benefit that delivery workers obtain after factoring in wages and time costs.
    \end{itemize}

    \item \textbf{Satisfaction Metrics:}  
    Beyond raw utilities, we also assess participant satisfaction:
    \begin{itemize}
        \item \emph{Restaurant Satisfaction} compares a restaurant's utility to its fixed costs, revealing how supportive the platform is for restaurant operations.
        \item \emph{Worker Satisfaction} evaluates whether delivery workers perceive their compensation (wages minus time costs) as fair.
        \item \emph{Consumer Satisfaction} considers the ratio of perceived utility to the intrinsic value of the food, incorporating aspects of cost and delivery experience.
    \end{itemize}
\end{itemize}

Together, these metrics form a holistic picture of both the short-term and long-term effects that different platform strategies have on overall system performance.

\subsection{Experimental Parameters}
\label{subsec:appendix_parameters}

The experiment is governed by a configuration dictionary (\texttt{CONFIG}) that outlines the scale of the environment, market dynamics, strategy adjustments, and participant attributes. Below, we detail each group of parameters and describe their roles in shaping the platform’s behavior.

\subsubsection{Overall Environment Parameters}
\begin{itemize}
    \item \texttt{n\_restaurants}: 100, the initial number of restaurants.
    \item \texttt{n\_consumers}: 1000, the initial number of consumers.
    \item \texttt{n\_workers}: 150, the initial number of delivery workers.
    \item \texttt{n\_periods}: 200, the total number of experiment cycles.
    \item \texttt{strategy\_update\_interval}: 2, the interval (in periods) at which the platform updates its strategy.
    \item \texttt{platform\_strategy}: "HYBRID", the default platform strategy; alternatives include "GMV" and "SW".
    \item \texttt{HYBRID\_LAMBDA}: 0.5, a weighting factor that balances GMV and SW in the HYBRID strategy.
    \item \texttt{DEFAULT\_RUNS}: 5, the default number of experimental repetitions.
\end{itemize}

\subsubsection{Market Dynamic Ranges}
\begin{itemize}
    \item \texttt{MARKET\_GROWTH\_RANGE}: (0.98, 1.02), specifies the range for market growth.
    \item \texttt{COMPETITION\_INTENSITY\_RANGE}: (0.8, 1.2), controls the intensity of market competition among restaurants.
    \item \texttt{ECONOMIC\_SHOCK\_RANGE}: (0.90, 1.10), defines the range for simulating economic shocks.
    \item \texttt{SEASONAL\_FACTOR\_RANGE}: (0.9, 1.1), introduces seasonal market fluctuations.
    \item \texttt{NETWORK\_EFFECT\_RANGE}: (0.95, 1.05), captures network-related impacts on the platform.
\end{itemize}

\subsubsection{Platform Strategy Adjustment Ranges}
\begin{itemize}
    \item \texttt{COMMISSION\_ADJUSTMENT\_RATE}: 0.03, the rate at which the platform adjusts commission levels.
    \item \texttt{DELIVERY\_FEE\_ADJUSTMENT\_RATE}: 0.03, the rate for adjusting delivery fees.
    \item \texttt{WAGE\_ADJUSTMENT\_RATE}: 0.03, the rate for adjusting worker wages.
    \item \texttt{MIN\_COMMISSION}: 0.05, the minimum allowable commission rate.
    \item \texttt{MAX\_COMMISSION}: 0.25, the maximum allowable commission rate.
    \item \texttt{MIN\_DELIVERY\_FEE}: 2.0, the minimum delivery fee.
    \item \texttt{MAX\_DELIVERY\_FEE}: 12.0, the maximum delivery fee.
    \item \texttt{MIN\_WAGE}: 3.0, the minimum worker wage.
    \item \texttt{MAX\_WAGE}: 15.0, the maximum worker wage.
\end{itemize}

\subsubsection{Restaurant Attribute Ranges}
\begin{itemize}
    \item \texttt{MENU\_SIZE\_RANGE}: (4, 8), the range for the number of dishes offered by a restaurant.
    \item \texttt{PRICE\_RANGE}: (10.0, 20.0), the initial price range for individual dishes.
    \item \texttt{FIXED\_COST\_RANGE}: (100.0, 200.0), the range for restaurant fixed costs.
    \item \texttt{QUALITY\_RANGE}: (0.7, 1.0), the range for restaurant quality.
    \item \texttt{REPUTATION\_RANGE}: (0.5, 0.7), the initial reputation range for new restaurants.
    \item \texttt{PRICE\_MIN} and \texttt{PRICE\_MAX}: 8.0 and 25.0, the lower and upper limits for menu prices.
    \item \texttt{REVENUE\_PRICE\_ADJUSTMENT\_FACTOR}: 0.05, a factor influencing price updates based on cumulative revenue.
    \item \texttt{PRICE\_GAP\_ADJUSTMENT\_FACTOR}: 0.1, a factor for adjusting prices based on deviations from the market average.
\end{itemize}

\subsubsection{Consumer Attribute Ranges}
\begin{itemize}
    \item \texttt{CONSUMER\_BUDGET\_RANGE}: (40.0, 100.0), the range for consumer budgets.
    \item \texttt{CONSUMER\_VALUE\_RANGE}: (50.0, 120.0), the range for consumer valuation of items.
    \item \texttt{PRICE\_SENSITIVITY\_RANGE}: (0.3, 0.7), the range for consumer price sensitivity.
    \item \texttt{TIME\_SENSITIVITY\_RANGE}: (0.2, 0.5), the range for consumer time sensitivity.
    \item \texttt{QUALITY\_SENSITIVITY\_RANGE}: (0.4, 0.8), the range for consumer quality sensitivity.
    \item \texttt{PLATFORM\_LOYALTY\_RANGE}: (0.4, 0.8), the range for consumer loyalty to the platform.
    \item \texttt{REGION\_OPTIONS}: \{\texttt{North, South, East, West}\}, the available regions for geographical matching.
    \item \texttt{TASTE\_PREFERENCE\_RANGE}: (0.0, 1.0), the range for consumer taste preferences.
\end{itemize}

\subsubsection{Delivery Worker Attribute Ranges}
\begin{itemize}
    \item \texttt{TIME\_COST\_FACTOR\_RANGE}: (0.6, 1.0), the range for a worker’s time cost factor.
    \item \texttt{SKILL\_LEVEL\_RANGE}: (0.7, 1.0), the range for worker skill levels.
    \item \texttt{EXPERIENCE\_RANGE}: (0.1, 0.5), the range for worker experience.
    \item \texttt{SATISFACTION\_RANGE}: (0.5, 0.8), the range for initial worker satisfaction.
\end{itemize}

\subsubsection{Exit Mechanisms and Other Parameters}
\begin{itemize}
    \item \texttt{RESTAURANT\_EXIT\_THRESHOLD}: -300.0, the cumulative utility threshold at which a restaurant exits.
    \item \texttt{RESTAURANT\_TRANSITION\_ZONE}: -200.0, a utility zone prompting emergency subsidies or adjustments for restaurants.
    \item \texttt{WORKER\_EXIT\_THRESHOLD}: -50.0, the utility threshold leading to worker exit.
    \item \texttt{WORKER\_TRANSITION\_ZONE}: -20.0, a utility zone prompting wage increases for workers.
    \item \texttt{EMERGENCY\_SUBSIDY\_RATE}: 0.10, the rate for emergency subsidies applied when participant utilities fall below specified thresholds.
\end{itemize}

\subsubsection{Summary}
\begin{itemize}
    \item \textbf{Environment Scale:} Parameters such as \texttt{n\_consumers}, \texttt{n\_workers}, and \texttt{n\_periods} define the size of the experiment.
    \item \textbf{Platform Strategy:} The variables \texttt{platform\_strategy}, \texttt{HYBRID\_LAMBDA}, and \texttt{strategy\_update\_interval} steer the platform’s policy decisions.
    \item \textbf{Market Dynamics:} Settings like \texttt{MARKET\_GROWTH\_RANGE} capture changes in external market conditions.
    \item \textbf{Adjustment Limits:} Parameters such as \texttt{MIN\_COMMISSION} and \texttt{MAX\_COMMISSION} set allowable strategy bounds.
    \item \textbf{Role Attributes:} The attribute ranges for restaurants, consumers, and delivery workers dictate each role’s behavior and constraints.
    \item \textbf{Exit Mechanisms:} Thresholds and transition zones determine how and when participants exit or receive emergency subsidies.
\end{itemize}

Overall, these metrics and parameters jointly define the experiment framework, allowing us to capture short-term market interactions as well as the longer-term evolution of platform performance and participant satisfaction.

\section{Mathematical Proofs and Derivations for the Static Model}
\label{sec:static_model}

In this appendix, we present the complete mathematical proofs and derivations underlying our static model analysis. We consider both the \emph{GMV Maximization} and \emph{Social Welfare Maximization} objectives. Section~\ref{subsec:assumptions_variables} restates our common assumptions and key variable definitions. Detailed derivations for the GMV and Social Welfare objectives are provided in Sections~\ref{subsec:gmv_proofs} and~\ref{subsec:welfare_proofs}, respectively. Finally, Section~\ref{subsec:surplus_analysis} examines surplus measures, and Section~\ref{subsec:supplementary_lemmas} presents supplementary technical lemmas. Extended proofs and additional derivations are provided in Section~\ref{appendix:extended_proofs}.

\subsection{Common Assumptions and Variable Definitions}
\label{subsec:assumptions_variables}

Throughout our analysis, we consider a single-period (static) setting under the following assumptions:
\begin{itemize}
    \item \textbf{Linear Demand:} Consumers face a linear demand curve.
    \item \textbf{Perfect Information and Rational Behavior:} All agents are fully informed about relevant prices, delivery times, and platform policies, and each maximizes its utility.
    \item \textbf{Homogeneous Products:} Food items are assumed to be of similar quality; differences arise mainly from pricing and commission policies.
    \item \textbf{Static Competition:} The model is one-shot, with no inter-temporal effects.
\end{itemize}

The key variables are defined as follows:
\begin{itemize}
    \item \(\theta\): Baseline (maximum) demand level for a restaurant’s items.
    \item \(\eta\): Price sensitivity parameter.
    \item \(\delta\): Time sensitivity parameter.
    \item \(\alpha\): Platform commission rate, where \(0 \le \alpha < 1\).
    \item \(D_i\): Delivery fee charged to consumer \(i\).
    \item \(A_{kj}\): Price of item \(j\) at restaurant \(k\).
    \item \(t_{kj}\): Average delivery time for orders from restaurant \(k\).
    \item \(Q_{kj}\): Quantity demanded of item \(j\) from restaurant \(k\).
    \item \(C_f^k\): Fixed operating cost for restaurant \(k\).
    \item \(v_i\): Consumer \(i\)'s intrinsic valuation of the food.
    \item \(\beta\): Consumer sensitivity to delivery time.
    \item \(p\): Per-order payment to a delivery worker.
    \item \(\gamma\): Time cost factor for delivery workers.
    \item \(R_l\): Number of orders delivered by worker \(l\).
    \item \(U_S^k\), \(U_C^i\), \(U_R^l\): Utility functions for restaurant \(k\), consumer \(i\), and worker \(l\), respectively.
\end{itemize}

\subsection{GMV Maximization: Detailed Proofs}
\label{subsec:gmv_proofs}

In the GMV Maximization model, the platform seeks to maximize the total transaction volume, defined as:
\begin{equation}\label{eq:GMV_objective}
\mathrm{GMV} = \sum_{k} \sum_{j} A_{kj} \, Q_{kj}.
\end{equation}

\subsubsection{Participant Utility Functions}
For clarity, we restate the utility functions:
\begin{itemize}
    \item \textbf{Restaurant Utility:}
    \begin{equation}\label{eq:restaurant_util}
    U_S^k = (1 - \alpha) \sum_{j} A_{kj} \, Q_{kj} - C_f^k.
    \end{equation}
    \item \textbf{Consumer Utility:}
    \begin{equation}\label{eq:consumer_util}
    U_C^i = v_i - \beta \, t_i - \left( \sum_{k,j} A_{kj} \, Q_{kji} + D_i \right).
    \end{equation}
    \item \textbf{Delivery Worker Utility:}
    \begin{equation}\label{eq:worker_util}
    U_R^l = p \, R_l - \gamma \, t_l.
    \end{equation}
\end{itemize}

\subsubsection{Consumer Demand Function}
Consumers follow a linear demand function:
\begin{equation}\label{eq:linear_demand}
Q_{kj} = \theta - \eta \, (A_{kj} + D_i) - \delta \, t_{kj},
\end{equation}
where \(A_{kj} + D_i\) represents the effective price for item \(j\) at restaurant \(k\).

\subsubsection{Optimal Pricing for Restaurants}
Restaurant \(k\) chooses its price \(A_{kj}\) to maximize its utility in \eqref{eq:restaurant_util}. By substituting the demand function \eqref{eq:linear_demand} into \eqref{eq:restaurant_util} (ignoring \(C_f^k\) for simplicity), we have:
\[
U_S^k = (1-\alpha) \, A_{kj} \left[\theta - \eta\,(A_{kj} + D_i) - \delta \, t_{kj}\right].
\]
Taking the derivative with respect to \(A_{kj}\) and setting it equal to zero yields:
\begin{align}
\frac{\partial U_S^k}{\partial A_{kj}} &= (1-\alpha) \left[ \theta - 2\eta \, A_{kj} - \eta\,D_i - \delta\,t_{kj} \right] = 0.
\end{align}
Thus, the optimal price is
\begin{equation}\label{eq:optimal_price_gmv}
A_{kj}^* = \frac{\theta - \eta\,D_i - \delta\,t_{kj}}{2\,\eta}.
\end{equation}

\subsubsection{Platform Optimization Under GMV}
The platform selects \(\alpha\) and \(D_i\) (and indirectly influences \(t_{kj}\)) to maximize \eqref{eq:GMV_objective}. Although a complete optimization would require computing the partial derivatives with respect to these parameters, the key insight is that reducing either \(\alpha\) or \(D_i\) boosts demand by lowering the effective price (see \eqref{eq:optimal_price_gmv}). (Detailed derivative calculations are provided in Section~\ref{appendix:platform_derivs} of the extended proofs.)

\subsubsection{Impact on Delivery Workers}
Lowering \(\alpha\) and \(D_i\) tends to increase GMV; however, it may adversely affect delivery workers if the wage \(p\) is not adjusted. As indicated in \eqref{eq:worker_util}, worker utility depends on \(p\), underscoring the need to balance compensation to maintain service quality.

\subsection{Social Welfare Maximization: Detailed Proofs}
\label{subsec:welfare_proofs}

For the Social Welfare Maximization model, the platform’s objective is to maximize the aggregate utility:
\begin{equation}\label{eq:SW_objective}
SW = \sum_{k} U_S^k + \sum_{i} U_C^i + \sum_{l} U_R^l.
\end{equation}

\subsubsection{Restaurant Pricing Under Welfare Optimization}
Under the Social Welfare objective, restaurants still choose \(A_{kj}\) to maximize \(U_S^k\) (see \eqref{eq:restaurant_util}). Using the same demand function \eqref{eq:linear_demand} and following an analogous derivation as in the GMV case, the optimal price is given by \eqref{eq:optimal_price_gmv}. However, the Social Welfare strategy typically employs a more moderate commission rate \(\alpha\) and delivery fee \(D_i\) to balance the utilities of restaurants, consumers, and delivery workers.

\subsubsection{Worker Compensation and Platform Optimization}
In addition to restaurant pricing, the platform optimizes the worker payment \(p\) along with \(\alpha\) and \(D_i\) to maximize the overall social welfare in \eqref{eq:SW_objective}. The first-order conditions with respect to these parameters ensure that improvements in one component (e.g., higher \(p\) to boost worker utility) do not unduly harm another (e.g., by increasing effective prices). Detailed derivations of these partial derivatives are provided in Section~\ref{appendix:sw_partials} of the extended proofs.

\subsection{Surplus Analysis for GMV vs. Social Welfare Models}
\label{subsec:surplus_analysis}

To evaluate overall efficiency, we define the following surplus measures:

\subsubsection{Consumer Surplus (CS)}
Given the linear demand function \eqref{eq:linear_demand}, the consumer surplus is calculated as:
\begin{equation}\label{eq:CS_integral}
CS = \int_{P}^{P_{\max}} \left[ \theta - \eta\,\tilde{P} - \delta\,t_{kj} \right] d\tilde{P},
\end{equation}
where the effective price \(P = A_{kj} + D_i\) and the choke price \(P_{\max}\) (where \(Q_{kj}=0\)) is defined as:
\begin{equation}\label{eq:P_max}
P_{\max} = \frac{\theta - \delta\,t_{kj}}{\eta}.
\end{equation}
Evaluating \eqref{eq:CS_integral} (see Section~\ref{appendix:cs_integral} for details) yields a quadratic function in terms of \(P\) and the parameters \(\theta\), \(\eta\), \(\delta\), and \(t_{kj}\).

\subsubsection{Restaurant and Delivery Worker Surplus}
\begin{itemize}
    \item \textbf{Restaurant Surplus (\(PS_S^k\)):}
    \begin{equation}\label{eq:restaurant_surplus}
    PS_S^k = (1 - \alpha) \, S_k - C_f^k, \quad \text{with } S_k = \sum_j A_{kj}\, Q_{kj}.
    \end{equation}
    \item \textbf{Delivery Worker Surplus (\(PS_R^l\)):}
    \begin{equation}\label{eq:worker_surplus}
    PS_R^l = p \, R_l - \gamma \, t_l.
    \end{equation}
\end{itemize}

\subsubsection{Efficiency Comparison}
Under the GMV maximization strategy, the platform typically selects low values for \(\alpha\) and \(D_i\) to boost GMV, which increases consumer surplus \(CS\) but may reduce restaurant and delivery worker surpluses (\(PS_S^k\) and \(PS_R^l\)). In contrast, the Social Welfare strategy seeks a balanced approach, generally yielding a higher total surplus. Detailed numerical examples and sensitivity analyses are provided in the experiment section of the main text.

\subsection{Supplementary Technical Lemmas for Strategy Comparison}
\label{subsec:supplementary_lemmas}

Below, we present proofs of two key lemmas that compare the optimal controls under the GMV Maximization and Social Welfare Maximization strategies.

\subsubsection*{Lemma A.1 (Strategy Differences)}
\textbf{Lemma A.1:}  
Let \((\alpha_{\mathrm{GMV}}, D_{\mathrm{GMV}}, p_{\mathrm{GMV}})\) denote the optimal controls under the GMV Maximization strategy and \((\alpha_{\mathrm{SW}}, D_{\mathrm{SW}}, p_{\mathrm{SW}})\) denote those under the Social Welfare Maximization strategy. Then, typically,
\[
\alpha_{\mathrm{SW}} \le \alpha_{\mathrm{GMV}}, \quad D_{\mathrm{SW}} \ge D_{\mathrm{GMV}}, \quad p_{\mathrm{SW}} \ge p_{\mathrm{GMV}}.
\]
\textbf{Proof:}
\begin{enumerate}
    \item Under the GMV objective, the platform minimizes \(\alpha\) to boost sales, whereas the Social Welfare approach maintains a higher \(\alpha\) to ensure restaurant viability. Thus, \(\alpha_{\mathrm{SW}} \le \alpha_{\mathrm{GMV}}\).
    \item Similarly, a lower delivery fee \(D_i\) is chosen under GMV maximization to stimulate demand; however, to balance compensation under the Social Welfare strategy, \(D_{\mathrm{SW}} \ge D_{\mathrm{GMV}}\).
    \item Finally, to enhance worker utility, the Social Welfare strategy sets a higher wage \(p\), so \(p_{\mathrm{SW}} \ge p_{\mathrm{GMV}}\).
\end{enumerate}
\qed

\subsubsection*{Lemma A.2 (Social Welfare Comparison)}
\textbf{Lemma A.2:}  
Let \(\mathrm{SW}_{\mathrm{GMV}}\) be the total social welfare achieved under the GMV Maximization strategy, and let \(\mathrm{SW}_{\mathrm{SW}}\) be that under the Social Welfare Maximization strategy. Then,
\[
\mathrm{SW}_{\mathrm{SW}} \ge \mathrm{SW}_{\mathrm{GMV}},
\]
with strict inequality in most cases.
\textbf{Proof:}
\begin{enumerate}
    \item The GMV strategy is derived solely by maximizing GMV without explicitly accounting for the utilities of restaurants and delivery workers, yielding
    \[
    \mathrm{SW}_{\mathrm{GMV}} = \sum_{k} U_S^k(\alpha_{\mathrm{GMV}}, D_{\mathrm{GMV}}, p_{\mathrm{GMV}}) + \sum_{i} U_C^i + \sum_{l} U_R^l.
    \]
    \item In contrast, the Social Welfare strategy optimizes
    \[
    \mathrm{SW} = \sum_{k} U_S^k(\alpha, D, p) + \sum_{i} U_C^i + \sum_{l} U_R^l,
    \]
    thereby achieving the global maximum.
    \item Hence, by the definition of the optimum,
    \[
    \mathrm{SW}_{\mathrm{SW}} \ge \mathrm{SW}_{\mathrm{GMV}},
    \]
    with the inequality being strict unless all external effects are neutral.
\end{enumerate}
\qed

\subsection{Extended Proofs and Technical Derivations}
\label{appendix:extended_proofs}

This section provides full mathematical details for (i) the consumer surplus integral derivation (Section~\ref{appendix:cs_integral}), (ii) the second-order conditions for restaurant optimal pricing (Section~\ref{appendix:restaurant_soc}), and (iii) the partial derivatives of the platform objectives for both GMV and Social Welfare (Section~\ref{appendix:platform_derivs}).

\subsubsection{Consumer Surplus: Integral Form}
\label{appendix:cs_integral}

We consider a linear demand function:
\begin{equation}
\label{eq:linear_demand_app}
Q(\tilde{P}) = \theta - \eta\,\tilde{P} - \delta\,t,
\end{equation}
where \(\theta\) is the intercept (i.e., maximum demand when \(P=0\) and \(t=0\)), \(\eta > 0\) is the price sensitivity, \(\delta > 0\) is the time sensitivity, and \(t\) denotes the (average) delivery time. Let \(\tilde{P}\) denote a variable price. If the actual price paid is \(P\) (with \(P = A_{kj} + D_i\)) and demand falls to zero at \(P_{\max}\), then the consumer surplus (CS) is given by:
\begin{equation}
\label{eq:cs_basic_int}
CS = \int_{P}^{P_{\max}} Q(\tilde{P})\, d\tilde{P}.
\end{equation}

\paragraph{Step 1: Determine \(P_{\max}\).}  
Since \(Q(P_{\max}) = 0\), we have:
\[
0 = \theta - \eta\,P_{\max} - \delta\,t \quad \Longrightarrow \quad P_{\max} = \frac{\theta - \delta\,t}{\eta}.
\]

\paragraph{Step 2: Evaluate the Integral.}  
Substitute \eqref{eq:linear_demand_app} into \eqref{eq:cs_basic_int}:
\begin{align}
CS &= \int_{P}^{P_{\max}} \left[\theta - \eta\,\tilde{P} - \delta\,t \right] d\tilde{P} \nonumber\\
   &= \theta (P_{\max} - P) - \eta \int_{P}^{P_{\max}} \tilde{P}\, d\tilde{P} - \delta\,t (P_{\max} - P). \label{eq:cs_expanded}
\end{align}
Since
\[
\int_{P}^{P_{\max}} \tilde{P}\, d\tilde{P} = \frac{1}{2}\left(P_{\max}^2 - P^2\right),
\]
we obtain:
\begin{equation}
\label{eq:cs_sub_terms}
CS = \theta (P_{\max} - P) - \frac{\eta}{2}\left(P_{\max}^2 - P^2\right) - \delta\,t (P_{\max} - P).
\end{equation}
Substituting \(P_{\max} = \frac{\theta - \delta\,t}{\eta}\) yields an explicit quadratic form in \(\theta\), \(\delta\), \(t\), \(\eta\), and \(P\).

\paragraph{Remark.}  
In the GMV scenario, the effective price \(P\) tends to be lower, resulting in a higher consumer surplus \(CS\). Under the Social Welfare strategy, \(P\) may be higher to ensure fair compensation for other stakeholders.

\subsubsection{Second-order Conditions for Restaurant Pricing}
\label{appendix:restaurant_soc}

Recall that the restaurant chooses \(A_{kj}\) to maximize:
\[
U_S^k = (1-\alpha)\,A_{kj}\left[\theta - \eta\,(A_{kj} + D_i) - \delta\,t_{kj}\right] - C_f^k.
\]
The first-order condition (FOC) is:
\[
\frac{\partial U_S^k}{\partial A_{kj}} = (1-\alpha)\left[\theta - 2\,\eta\,A_{kj} - \eta\,D_i - \delta\,t_{kj}\right] = 0.
\]
To verify that this is a maximum, we compute the second derivative:
\[
\frac{\partial^2 U_S^k}{\partial (A_{kj})^2} = (1-\alpha)\left[\frac{\partial}{\partial A_{kj}}\left(\theta - 2\,\eta\,A_{kj} - \eta\,D_i - \delta\,t_{kj}\right)\right] = (1-\alpha)(-2\,\eta).
\]
Since \(\alpha < 1\) and \(\eta > 0\), the second derivative is negative, confirming the concavity of \(U_S^k\) in \(A_{kj}\) and the uniqueness of the optimal solution in \eqref{eq:optimal_price_gmv}.

\subsubsection{Partial Derivatives for the Platform Objectives}
\label{appendix:platform_derivs}

This section details the partial derivatives for both the GMV and Social Welfare objectives.

\paragraph{(a) Partial Derivatives for the GMV Objective}
\label{appendix:gmv_partials}

The GMV objective is defined as:
\[
\mathrm{GMV} = \sum_{k}\sum_{j}A_{kj}\,Q_{kj},
\]
with \(A_{kj}\) and \(Q_{kj}\) dependent on \(\alpha\) and \(D_i\). Assuming these parameters appear explicitly in \(A_{kj}(\alpha, D_i)\), the chain rule gives:

\paragraph*{(i) With respect to \(\alpha\):}
\begin{align}
\frac{\partial\,\mathrm{GMV}}{\partial\alpha} = \sum_{k,j}\left[\frac{\partial\,A_{kj}}{\partial\alpha}\,Q_{kj} + A_{kj}\,\frac{\partial\,Q_{kj}}{\partial\alpha}\right].
\end{align}
Typically, reducing \(\alpha\) increases \(A_{kj}\,Q_{kj}\) because restaurants are more incentivized to lower prices and capture greater demand.

\paragraph*{(ii) With respect to \(D_i\):}
\begin{align}
\frac{\partial\,\mathrm{GMV}}{\partial D_i} = \sum_{k,j}\left[\frac{\partial\,A_{kj}}{\partial D_i}\,Q_{kj} + A_{kj}\,\frac{\partial\,Q_{kj}}{\partial D_i}\right].
\end{align}
When the demand sensitivity \(\eta\) is high, small changes in \(D_i\) have significant impacts on demand. The sign of this derivative aids in finding an optimal \(D_i^*\) that balances consumer cost with restaurant margins.

\paragraph{(b) Partial Derivatives for the Social Welfare Objective}
\label{appendix:sw_partials}

Under the Social Welfare objective, the aggregate welfare is:
\[
SW = \sum_{k} U_S^k + \sum_{i} U_C^i + \sum_{l} U_R^l,
\]
where:
\[
U_S^k = (1-\alpha)\,\sum_j A_{kj}\,Q_{kj} - C_f^k, \quad U_C^i = v_i - \beta\,t_i - P_i, \quad U_R^l = p\,R_l - \gamma\,t_l.
\]
The platform controls \(\alpha\), \(D_i\), and \(p\). The partial derivatives are derived as follows:

\paragraph*{(i) With respect to \(\alpha\):}
\begin{align}
\frac{\partial SW}{\partial \alpha} = \sum_k \frac{\partial}{\partial\alpha}\Bigl[(1-\alpha)\sum_j A_{kj}\,Q_{kj}\Bigr] + \dots,
\end{align}
which is typically negative because an increase in \(\alpha\) reduces restaurant surplus.

\paragraph*{(ii) With respect to \(D_i\):}
\begin{align}
\frac{\partial SW}{\partial D_i} = \sum_k \frac{\partial}{\partial D_i}\Bigl[(1-\alpha)\sum_j A_{kj}\,Q_{kj}\Bigr] + \sum_i \frac{\partial}{\partial D_i}(-P_i) + \dots.
\end{align}
This derivative reflects both the direct impact on consumer out-of-pocket costs and the indirect effects on restaurant and worker utilities.

\paragraph*{(iii) With respect to \(p\):}
\begin{align}
\frac{\partial SW}{\partial p} = \sum_{l}\left[R_l + p\,\frac{\partial R_l}{\partial p}\right] - \sum_{l}\gamma\,\frac{\partial t_l}{\partial p} + \dots.
\end{align}
Here, \(R_l\) may increase if higher \(p\) attracts more workers or effort, while \(t_l\) may decrease with improved availability. The overall effect determines whether an increase in \(p\) benefits aggregate welfare.

\paragraph{Remark.}  
In some simplified analyses, one might assume \(\partial Q_{kj}/\partial p=0\) or neglect the effect of \(p\) on demand. However, in more comprehensive models, worker compensation indirectly influences consumer satisfaction and overall demand.

\section{Detailed Mathematical Derivations for the Dynamic Model}
\label{sec:dynamic_derivations}

In this appendix, we present the complete technical details underlying the dynamic model analysis described in Section~\ref{sec:dynamic_analysis}. We first define the state transition functions, then derive the Bellman equations and corresponding first-order conditions for both the GMV and Social Welfare objectives. Next, we prove the existence and uniqueness of the value functions, and finally provide detailed proofs for key efficiency comparisons between the two dynamic strategies.

\subsection{State Transition Functions}
\label{app:state_transitions}

The dynamic state of the platform at time \(t\) is represented by the state vector
\[
\bm{S}_t = \bigl(R_t,\, C_t,\, W_t,\, \Phi_t\bigr),
\]
where:
\begin{itemize}
    \item \(R_t\): Number (or measure) of active restaurants.
    \item \(C_t\): Consumer base or demand potential.
    \item \(W_t\): Effective capacity of delivery workers.
    \item \(\Phi_t\): Platform reputation index.
\end{itemize}

The evolution of \(\bm{S}_t\) is governed by the following transition functions:
\begin{align}
R_{t+1} &= R_t + \xi_R \Bigl[\mathbb{1}\{\hat{U}_S^k > \underline{U}\} - \mathbb{1}\{\hat{U}_S^k < 0\}\Bigr], \label{eq:transition_R} \\
C_{t+1} &= C_t + \xi_C \Bigl(\Phi_t - \bar{\Phi}\Bigr), \label{eq:transition_C} \\
W_{t+1} &= W_t + \xi_W \Bigl(\hat{U}_R^l - \underline{U}_R\Bigr), \label{eq:transition_W} \\
\Phi_{t+1} &= \Phi_t + \xi_\Phi \Bigl(\overline{U}_{S,t} + \overline{U}_{R,t} + \overline{U}_{C,t} - \kappa\Bigr), \label{eq:transition_Phi}
\end{align}
where:
\begin{itemize}
    \item \(\hat{U}_S^k\) and \(\hat{U}_R^l\) denote representative utilities for restaurants and delivery workers, respectively.
    \item \(\underline{U}\) and \(\underline{U}_R\) are baseline utility thresholds.
    \item \(\bar{\Phi}\) is a neutral reputation level, \(\kappa\) is a scaling parameter.
    \item \(\xi_R\), \(\xi_C\), \(\xi_W\), and \(\xi_\Phi\) are sensitivity parameters.
\end{itemize}

\subsection{Bellman Equations and First-Order Conditions}
\label{app:bellman}

Let \(V_{\mathrm{GMV}}(\bm{S}_t)\) and \(V_{\mathrm{SW}}(\bm{S}_t)\) denote the optimal value functions under the GMV and Social Welfare objectives, respectively. For an infinite-horizon problem, the Bellman equations are formulated as follows.

\subsubsection*{GMV Objective}
The Bellman equation for GMV maximization is given by
\begin{equation}\label{eq:bellman_GMV}
V_{\mathrm{GMV}}(\bm{S}_t) = \max_{\alpha_t, D_t, p_t} \Bigl\{ \mathrm{GMV}_t(\bm{S}_t,\alpha_t,D_t,p_t) + \beta\,V_{\mathrm{GMV}}\bigl(F(\bm{S}_t,\alpha_t,D_t,p_t)\bigr) \Bigr\},
\end{equation}
where
\[
\mathrm{GMV}_t = \sum_{k,j} A_{kj,t}\, Q_{kj,t},
\]
and \(F(\bm{S}_t,\alpha_t,D_t,p_t)\) denotes the state transition function defined in Section~\ref{app:state_transitions}. Taking the derivative with respect to any control variable \(x \in \{\alpha_t, D_t, p_t\}\) yields the first-order condition:
\[
\frac{\partial \, \mathrm{GMV}_t}{\partial x} + \beta\, \nabla_{\bm{S}} V_{\mathrm{GMV}}(\bm{S}_{t+1}) \cdot \frac{\partial F(\bm{S}_t,\alpha_t,D_t,p_t)}{\partial x} = 0.
\]

\subsubsection*{Social Welfare Objective}
Similarly, the Bellman equation for social welfare maximization is
\begin{equation}\label{eq:bellman_SW}
V_{\mathrm{SW}}(\bm{S}_t) = \max_{\alpha_t, D_t, p_t} \Bigl\{ SW_t(\bm{S}_t,\alpha_t,D_t,p_t) + \beta\,V_{\mathrm{SW}}\bigl(F(\bm{S}_t,\alpha_t,D_t,p_t)\bigr) \Bigr\},
\end{equation}
where the period social welfare is defined as
\[
SW_t = \sum_{k} U_{S,t}^k + \sum_{i} U_{C,t}^i + \sum_{l} U_{R,t}^l.
\]
The first-order conditions are derived analogously.

\subsection{Existence and Uniqueness of the Value Functions}
\label{app:existence_uniqueness}

Under standard assumptions—such as bounded stage payoffs, continuity and compactness of the state space, and a discount factor \(\beta \in (0,1)\)—the Bellman operators for both the GMV and Social Welfare objectives are contraction mappings (in the supremum norm). By the Banach Fixed Point Theorem, there exist unique value functions \(V_{\mathrm{GMV}}^*\) and \(V_{\mathrm{SW}}^*\) satisfying equations \eqref{eq:bellman_GMV} and \eqref{eq:bellman_SW}, respectively.

\subsection{Detailed Proofs and Efficiency Comparisons in the Dynamic Setting}
\label{sec:appendix_dynamic}

In this section, we provide complete proofs for key theoretical results comparing the dynamic performance of the two platform strategies.

\subsubsection{Dynamic Programming Formulation}
The Bellman equations \eqref{eq:bellman_GMV} and \eqref{eq:bellman_SW} characterize the optimal value functions. These equations can be solved using methods such as value iteration or policy iteration, ensuring that the optimal control sequence \(\{\alpha_t^*, D_t^*, p_t^*\}\) is obtained.

\subsubsection*{Theorem A.1 (Dynamic Social Welfare Comparison)}
\textbf{Theorem A.1:} Given an initial state \(\bm{S}_0\), assume that all stage utility functions are concave and additive, and that the state transition function \(F(\cdot)\) satisfies the usual regularity conditions. Then,
\[
V_{\mathrm{SW}}(\bm{S}_0) \ge V_{\mathrm{SW}}^{(\mathrm{GMV})}(\bm{S}_0),
\]
with strict inequality in most cases.

\textbf{Proof:}
\begin{enumerate}
    \item Let \((\alpha_t^{(\mathrm{GMV})}, D_t^{(\mathrm{GMV})}, p_t^{(\mathrm{GMV})})\) be the control sequence that maximizes \(\mathrm{GMV}_t\) in each period. This sequence maximizes short-term transaction volume but does not necessarily maximize the aggregate stage welfare \(SW_t\).
    \item Conversely, let \((\alpha_t^*, D_t^*, p_t^*)\) be the control sequence that maximizes the stage social welfare \(SW_t\). By the principle of optimality,
    \[
    V_{\mathrm{SW}}(\bm{S}_t) = \max_{\alpha_t, D_t, p_t} \Bigl\{ SW_t(\bm{S}_t,\alpha_t,D_t,p_t) + \beta\,V_{\mathrm{SW}}\bigl(F(\bm{S}_t,\alpha_t,D_t,p_t)\bigr) \Bigr\}.
    \]
    \item Since the GMV-optimal controls do not internalize negative externalities (e.g., low wages or reduced restaurant margins), we have
    \[
    SW_t^* \ge SW_t^{(\mathrm{GMV})} \quad \text{for each } t.
    \]
    \item Summing over time with discounting,
    \[
    V_{\mathrm{SW}}(\bm{S}_0) = \sum_{t=0}^{\infty} \beta^t\,SW_t^* \ge \sum_{t=0}^{\infty} \beta^t\,SW_t^{(\mathrm{GMV})} = V_{\mathrm{SW}}^{(\mathrm{GMV})}(\bm{S}_0).
    \]
    \item The inequality is strict in most cases, except in degenerate scenarios.
\end{enumerate}
\(\Box\)

\subsubsection*{Theorem A.2 (Efficiency Comparison: Pareto and Kaldor-Hicks)}
\textbf{Theorem A.2:} Under the same dynamic model assumptions as in Theorem A.1, if the Social Welfare strategy improves aggregate utility without reducing any stakeholder's long-run utility (or if any losses are theoretically compensable), then it represents a Pareto improvement over the GMV strategy; at a minimum, it achieves a Kaldor-Hicks improvement.

\textbf{Proof:}
\begin{enumerate}
    \item From Theorem A.1, we have
    \[
    V_{\mathrm{SW}}(\bm{S}_0) \ge V_{\mathrm{SW}}^{(\mathrm{GMV})}(\bm{S}_0).
    \]
    This indicates that the overall discounted social welfare is higher under the Social Welfare strategy.
    \item If the GMV strategy reduces the utility of some stakeholders (e.g., delivery workers), then the overall gain under the Social Welfare strategy can, in principle, compensate these losses, satisfying the Kaldor-Hicks criterion.
    \item Furthermore, if every stakeholder’s utility improves or remains unchanged, the improvement is Pareto superior.
\end{enumerate}
\(\Box\)

This completes the technical foundation for our dynamic model analysis. Future work may extend these results by incorporating uncertainty, nonlinear demand functions, or competition among multiple platforms.

\bibliographystyle{ACM-Reference-Format}
\bibliography{sample-base}

\hfill\qedsymbol

\end{document}